\documentclass{ws-rv9x6}
\usepackage{subfigure} 
\usepackage{ws-rv-van}
\usepackage{mathrsfs}
\makeindex
\newcommand{\eg}{{\it e.g.}}
\newcommand{\ie}{{\it i.e.}}

\begin{document}

\chapter[First passage phenomena in bounded
domains]{Trajectory-to-trajectory fluctuations in first-passage
  phenomena in bounded domains}
\label{ra_ch1}

\author[T. G. Mattos]{Thiago G. Mattos}
%\index[aindx]{Mattos, T. G.}
\address{Max Planck Institute for Intelligent Systems, 
70569 Stuttgart, Germany. tgmattos@is.mpg.de}

\author[C. Mej\'ia-Monasterio]{Carlos Mej\'ia-Monasterio}
%\index[aindx]{Mej\'ia-Monasterio, C.}
\address{Department of Mathematics and
  Statistics, University of Helsinki, 00014 Helsinki, Finland \&  Laboratory of Physical Properties,
Technical University of Madrid, 28040 Madrid, Spain.
carlos.mejia@helsinki.fi}

\author[R. Metzler]{Ralf Metzler}
%\index[aindx]{Metzler, R.}
\address{Institute for Physics \& Astronomy, University of Postdam, 14476
Potsdam-Golm, Germany \& Physics Department,  Tampere University of
Technology, 33101 Tampere, Finland. rmetzler@uni-potsdam.de}

\author[G. Oshanin]{Gleb Oshanin}
%\index[aindx]{Oshanin, G.}
\address{CNRS -- Universit{\'e} Pierre et Marie Curie, LPTMC  UMR7600, 75252 Paris, France. oshanin@lptmc.jussieu.fr}

\author[T. G. Mattos, C. Mej\'ia-Monasterio, R. Metzler, G. Oshanin \&
G. Schehr]{Gr\'egory Schehr}
%\index[aindx]{Schehr, G.}
\address{CNRS -- Universit\'e Paris-Sud, LPTMS, 91405 Orsay, France
gregory.schehr@u-psud.fr}

\begin{abstract}
  We study the  statistics of the first passage of  a random walker to
  absorbing subsets  of the boundary  of compact domains  in different
  spatial  dimensions.   We  describe  a novel  diagnostic  method  to
  quantify  the  trajectory-to-trajectory  fluctuations of  the  first
  passage, based on the distribution of the so-called uniformity index
  $\omega$, measuring the similarity of the first passage times of two
  independent walkers starting at the same location.  We show that the
  characteristic  shape  of  $P(\omega)$  exhibits a  transition  from
  unimodal  to  bimodal,  depending  on  the  starting  point  of  the
  trajectories.  From  the study of  different geometries in  one, two
  and three dimensions, we conclude  that this transition is a generic
  property of first passage phenomena in bounded domains.  Our results
  show  that, in  general, the  Mean First  Passage Time  (MFPT)  is a
  meaningful  characteristic measure  of the  first  passage behaviour
  only  when the  Brownian  walkers start  sufficiently  far from  the
  absorbing  boundary.  Strikingly,  in the  opposite case,  the first
  passage    statistics    exhibit   large    trajectory-to-trajectory
  fluctuations  and  the MFPT  is  not  representative  of the  actual
  behaviour.
\end{abstract}
%\markright{Customized Running Head for Odd Page} % default is chapter title.
\body

\section{Introduction}
\label{sec:intro}

The concept  of first  passage underlies diverse  stochastic phenomena
for which the crucial aspect is  an event when some random variable of
interest  reaches a  preset value  for the  first time.   A few  stray
examples    across    disciplines     include    chemical    reactions
\cite{ol,lov,ben,beni,mattosreis,smolu},  the   firing  of   a  neuron
\cite{gersh,burkitt}, random search of a  mobile or an immobile target
\cite{alb,alb1,ol1a,ol1b,ol1c,ralfa,ralfb,ralfc,osh1,osh2,osh3,wio1,wio2,carlos,thiago,paul,port,evans1,evans2,gel},
diffusional disease  spreading \cite{Lloyd2001}, DNA  bubble breathing
\cite{hanke1,hanke2},      dynamics      of      molecular      motors
\cite{motor1a,motor1b,motor2},  the  triggering   of  a  stock  option
\cite{bouch}, etc. One distinguishes also between continuously varying
variables  for which  the first  passage across  a given  preset value
coincides  with  the   first  arrival  to  exactly   this  value,  and
discontinuous processes  for which this  value can be overshoot,  as it
happens, \eg,  for L{\'e}vy flights characterised  by long-tailed jump
length distributions with  diverging variance \cite{checha,chechb}.  A
variety of first passage time  phenomena and different related results
have  been   presented  in  Refs.~\refcite{katja,sid}.    This  volume
presents an  even broader exposition  of relevant areas  and describes
the state  of the  art in physical  and mathematical  comprehension of
such phenomena.

In this chapter we will be  concerned with first passage phenomena for
particles  executing  Brownian  motion.   The  distribution  of  first
passage times  for such  processes occurring  in unbounded  domains is
typically broad, such that not even the mean first passage time exists
\cite{sid}.  In particular,  in one-dimensional, semi-infinite domains
the  first  passage  time  distribution  of  a  Markovian  process  is
universally dominated  by the  $t^{-3/2}$ scaling  nailed down  by the
Sparre Andersen theorem \cite{sid}.  A  similar divergence of the mean
first  passage time  occurs in  stochastic processes  characterised by
scale-free distributions  of waiting times  \cite{mekla1,mekla2}. This
signifies    that    sample-to-sample   or    trajectory-to-trajectory
fluctuations are  of a crucial  importance for such processes  so that
the mean values --- the mean  first passage times, even if they exist
--- have little physical meaning.

However,  in  many  practically  important  situations  first  passage
processes involve  Brownian particles moving in  bounded domains (see,
\eg,  Refs.~\refcite{ol2,ol3a,ol3b,ol4a,ol4b,fBM-wedge}).  In  this case
the random variable  of interest, \eg, the first  passage time $\tau$
to a boundary, a target chemical  group, a binding site on the surface
of the domain or elsewhere within the domain, etc., has a distribution
$\Psi(\tau)$  which  possesses  moments  of  arbitrary,  negative  and
positive   order.    Such   distributions   are   usually   considered
\textit{narrow}, as  opposed to \textit{broad\/}  distributions, which
do            not            possess            all            moments
\cite{katja,sid,mekla1,mekla2,checha,chechb}.

Nonetheless, even  for such  distributions first passage  variables do
fluctuate, of  course, from sample to sample  or from trajectory  to 
trajectory,  and  an  important  question  is  how  to  quantify  such
fluctuations.   One  usually  resorts  to  standard  measures  of  the
statistical analysis, such as, \eg, standard deviations, the skewness
or the  kurtosis of the  distribution. But these quantities  are often
not very  instructive since they just  produce numbers and  it is not
always evident how to interpret or compare them.  Clearly, quantifying
the trajectory-to-trajectory  fluctuations in a  robust and meaningful
way is a challenging problem with many important applications.

In this  chapter we  focus on this  non-trivial problem and  discuss a
recently proposed procedure which presents a lucid illustration of the
effect  of the  trajectory-to-trajectory fluctuations.   It  allows to
quantify  their impact  and, moreover,  shows that  in some  cases the
distributions  considered  as  \textit{narrow} behave  effectively  as
broad   distributions  \cite{carlos,thiago}.    We   employ  a   novel
diagnostic  method based  on the  concept of  \textit{simultaneity} of
first passage events. Instead of the original first passage problem of
quantifying the  statistical outcome for a single  Brownian walker, in
this procedure one  simultaneously launches two identical, independent
Brownian  particles  at the  same  position  ${\bf  r_0}$, namely  two
different realisations  of a single  Brownian Motion (BM)  starting at
${\bf r_0}$.  The corresponding  outcomes are the first passage times
$\tau_1$ and $\tau_2$. One defines then the random variable
\begin{equation}
\label{def:omega}
\omega\equiv\frac{\tau_1}{\tau_1+\tau_2},
\end{equation}
such   that   $\omega$   ranges   in  the   interval   $[0,1]$.    The
\textit{uniformity      index\/}       $\omega$      measures      the
\textit{likelihood\/}  that   both  walkers   arrive  to   the  target
simultaneously: when $\omega$ is close  to 1/2, the process is uniform
and the particles behave as if  they were almost performing a Prussian
\textit{Gleichschritt}. In contrast, values of  $\omega$ close to 0 or
1   mean  highly   non-uniform   behaviour,   strongly  affecting   the
trajectory-to-trajectory fluctuations.   We note  parenthetically that
similar  random variables  have been  used in  the analysis  of random
probabilities   induced  by   normalisation  of   self-similar  L\'evy
processes~\cite{iddo1},  of the  fractal characterisation  of Paretian
Poisson  processes~\cite{iddo2},  and  of  the  so-called  Matchmaking
paradox~\cite{iddo3,iddo4},   and    more   recently,    to   quantify
sample-to-sample         fluctuations          in         mathematical
finances~\cite{samor2,hol},    chaotic   systems~\cite{schehr},    the
analysis  of distributions  of the  diffusion coefficient  of proteins
diffusing  along  DNAs\cite{boyer1}  and  optimal  estimators  of  the
diffusion coefficient of a single Brownian trajectory.\cite{boyer2}

To  illustrate  the  concept  of  simultaneity,  consider  a  generic,
\emph{generalized inverse Gaussian} form of $\Psi(\tau)$ (see \eg, the
discussion in Ref.~\refcite{carlos,thiago} and references therein)
\begin{equation}
\label{eq:Pt-exptrunc}
\Psi(\tau)\sim \exp\left(-\frac{a}{\tau}\right)\frac{1}{\tau^{1+\mu}}\exp
\left(-\frac{\tau}{b}\right),
\end{equation}
where $a$ and $b$ are some constants. Note that the first passage time
distribution in Eq.~(\ref{eq:Pt-exptrunc})  is exact in the particular
case of Brownian motion on a  semi-infinite line in presence of a bias
pointing towards the target site, or, equivalently, for the celebrated
integrate-and-fire model  of neuron firing by  Gerstein and Mandelbrot
\cite{gersh}.  In  general,  the  detailed form  of  $\Psi(\tau)$  for
diffusion in bounded domains is obviously much more complex than given
by  Eq.~(\ref{eq:Pt-exptrunc}), depending  on  the  shape of  the
domain under consideration and the exact boundary value problem and is
typically given in terms of an infinite series.

Nonetheless, on a  \textit{qualitative\/} level \eref{eq:Pt-exptrunc}
provides a clear  picture of the actual behaviour of  the first passage
time distribution in bounded domains. Namely, $\Psi(\tau)$ consists of
three different  parts: a singular  decay for small values  of $\tau$,
which mirrors the  fact that the first passage to  some point starting
from a distant position cannot occur instantaneously. This is followed
at intermediate times by a generic power-law decay with exponent $\mu$
--- the  so-called persistence exponent  \cite{satya,persist}, depending
on the exact type of random motion and spatial dimension.  Finally, an
exponential decay  at long $\tau$  cuts off the power-law.   A crucial
aspect is that  the exponential cutoffs at both short  and long $\tau$
ensure  that  in bounded  domains  $\Psi(\tau)$  possesses moments  of
arbitrary positive or  negative order. The parameters $a$  and $b$ 
depend on the  shape  of the  domain, its  typical  size, and  the
starting position of the particle  within the domain.  When the linear
size of the domain  (say, the radius $R$ of a  circular or a spherical
domain) diverges (\ie, $R\to\infty$), the parameter $b$ also diverges
such that the long-time asymptotic behaviour of the first passage time
distribution is of power-law form without  a cutoff.  In this case, at
least some, if not all, of the moments of $\Psi(\tau)$ diverge.

Now,  given the  distribution  $\Psi(\tau)$, the  distribution of  the
uniformity  index $\omega$  in  Eq. (\ref{def:omega})  can be  readily
calculated to give \cite{hol,schehr}
\begin{equation}
\label{def:Pw}
P(\omega) = \frac{1}{(1-\omega)^2} \int^{\infty}_0 \tau d\tau \Psi(\tau) \Psi\left(\frac{\omega}{1 - \omega} \tau\right) \,.
\end{equation}
As we explicitly show in the following section~\ref{sec:FPT}, for the
generic  distribution in  Eq.   (\ref{eq:Pt-exptrunc})  one finds  the
following explicit form \cite{thiago}:
\begin{equation}
\label{eq:Pw-exptrunc} 
P(\omega) = \frac{1}{2 K_{\mu}^2\left(2 \sqrt{a/b}\right)} \frac{1}{\omega (1- \omega)} K_{2 \mu}\left(2 \sqrt{\frac{a}{b \omega (1 - \omega)}}\right) \,,
 \end{equation}
 where  $K_{2 \mu}(\cdot)$  is  the modified  Bessel  function of  the
 second type.

 One notices  first that the  form of the distribution  $P(\omega)$ in
 Eq.~(\ref{eq:Pw-exptrunc})  is distinctly sensitive  to the  value of
 the  persistence  exponent  $\mu$,  which characterises  the  scaling
 behaviour  of  the first  passage  time  distribution $\Psi(\tau)$  at
 intermediate  times.  Thus,  for  $\mu>1$, $P(\omega)$  is  always  a
 unimodal, bell-shaped function with  a maximum at $\omega=1/2$, which
 implies  that in  such  a  situation both  walkers  (or two  distinct
 trajectories of one walker) will most likely arrive to the  target
 simultaneously.   For  $\mu=1$,   $P(\omega)$   is  almost   uniform,
 $P(\omega)\approx1$, apart from narrow regions at the corners $\omega
 =0$  and $\omega=1$,  for  $b/a\gg1$. Curiously,  for $\mu<1$,  which
 corresponds to  the most common  case, there exists a  critical value
 $p_c$ of  the ratio  $p=b/a$ such that  for $p>p_c$  the distribution
 $P(\omega)$ has a characteristic  M-shaped form with two maxima close
 to 0 and  1, while at $\omega=1/2$ one finds a  local minimum. Such a
 transition  from a  unimodal, bell-shaped  to bimodal,  M-shaped form
 mirrors a significant manifestation of sample-to-sample fluctuations.

 In  what follows  we survey  known results  and further  explore this
 intriguing behaviour of the first passage time distribution and of the
 corresponding  distribution  of  the  uniformity index  $\omega$  for
 diffusion in  bounded domains, focusing  on the effects of  the domain
 shape,  dimensionality,  location of  the  target,  the  type of  the
 boundary conditions, and of the initial position of the walker.

\section{First Passage Times and the Uniformity Distribution}
\label{sec:FPT}

Consider a  BM inside a general  $d$-dimensional domain $\mathcal{S}$,
whose                                                         boundary
$\partial\mathcal{S}\equiv\partial\mathcal{S}_a\cup\partial\mathcal{
  S}_r$ comprises reflecting,  $\partial\mathcal{S}_r$, and absorbing,
$\partial \mathcal{S}_a$,  parts. At time  $t=0$, the BM  initiates at
$\mathbf{r}_0\in \mathcal{S}$ and evolves  within the domain until the
trajectory hits  $\partial \mathcal{S}_a$ for  the first time  at some
random instant $\tau$. Furthermore, let $P(\mathbf{r},t|\mathbf{r}_0)$
denote  the  conditional  probability  distribution  for  finding  the
Brownian walker  at position  $\mathbf{r}$ at  time $t$,  provided the
initial condition was at $\mathbf{r}_0$ at $t=0$. The distribution $P(
\mathbf{r},t|\mathbf{r}_0)$ is the solution of the diffusion equation
\begin{equation}
\label{diff}
{\partial_t}P(\mathbf{r},t|\mathbf{r}_0)=D~ \partial^2_{\mathbf{r}}
P(\mathbf{r},t|\mathbf{r}_0)
\end{equation}
on  $\mathcal{S}$, subject  to the  initial condition  as well  as the
boundary  conditions   at  $\partial\mathcal{S}$.  Here   $D$  is  the
diffusion    coefficient   and   $\partial^2_{\mathbf{r}}$    is   the
$d$-dimensional  Laplacian. The solution  of this
initial boundary value  problem is, in the best  case, cumbersome, and
explicit  solutions may be  obtained for  only few  simple geometries,
(see \eg Ref.~\refcite{Carslaw}).

If   a   finite   part   of   the   boundary   is   absorbing,   \ie,
$\partial\mathcal{S}_a$   is   not   empty,  then   the   distribution
$P(\mathbf{r},t|\mathbf{r}_0)$ is no  longer normalised.  The survival
probability  $\mathscr{S}_{\mathbf{r}_0}(t)$ that  the walker  has not
reached $\partial\mathcal{S}_a$ up to time $t$, is defined by
\begin{equation}
\label{surv}
\mathscr{S}_{\mathbf{r}_0}(t)=\int_{\mathcal{S}}P(\mathbf{r},t|\mathbf{r}_0)
d\mathbf{r}.
\end{equation}
$\mathscr{S}_{\mathbf{r}_0}(t)$ is a monotonically decreasing function
of       time,       eventually       reaching       zero       value,
$\lim_{t\to\infty}\mathscr{S}_{\mathbf{r}  _0}(t)=0$. In terms  of the
survival probability,  the distribution of first passage  times to the
absorbing boundary becomes
\begin{equation}
\label{FPT}
\Psi_{\mathbf{r}_0}(\tau)=-\frac{d\mathscr{S}_{\mathbf{r}_0}(\tau)}{d\tau}
\ ,
\end{equation}
and the MFPT associated  with the distribution $\Psi(\tau)$ is defined
as the first moment
\begin{equation}
\label{MFPT}
\langle\tau\rangle(\mathbf{r}_0)=\int_0^\infty\tau\Psi_{\mathbf{r}_0}(\tau)
d\tau=\int_0^\infty\mathscr{S}_{\mathbf{r}_0}(\tau)d\tau.
\end{equation}
In most of the existing literature,  the dependence of the MFPT on the
starting position of  the walker is either simply  neglected, or it is
assumed  that the starting  point is  randomly distributed  within the
domain   $\mathcal{S}$.   However,   as  we   proceed  to   show,  the
$\mathbf{r}_0$-dependence of the first  passage time distribution is a
crucial aspect which cannot be neglected.

From the First Passage Time Distribution (FPTD) one readily obtains the
distribution of the uniformity  index $\omega$ defined in \eref{def:omega}, for
two independent BM, in terms of its moment generating function
\begin{equation}
\label{mg}
\Phi(\lambda)=\int_0^1P(\omega)\exp\left(-\lambda\omega\right)d\omega,
\end{equation}
with  $\lambda\geq0$.  Since  $\tau_1$ and  $\tau_2$  are independent,
identically  distributed random  variables, expression  (\ref{mg}) can be
formally represented as
\begin{equation}
\label{2}
\Phi(\lambda)=\int^{\infty}_0\int^{\infty}_0\Psi(\tau_1)\Psi(\tau_2)\exp\left(
-\lambda\frac{\tau_1}{\tau_1+\tau_2}\right)d\tau_1d\tau_2.
\end{equation}
Integrating over $d\tau_1$ we change the integration variable, $\tau_1\to
\omega$, so that Eq.~(\ref{2}) is rewritten in the form
\begin{equation}
\label{3}
\Phi(\lambda)=\int^{1}_0\exp\left(-\lambda\omega\right)\frac{d\omega}{(1-
\omega)^2}\int^{\infty}_0\tau_2\Psi(\tau_2)\Psi\left(\frac{\omega}{1-\omega}\tau_2\right)
d\tau_2.
\end{equation}
From  comparison  with  Eq.~(\ref{mg}),  one  obtains  $P(\omega)$  in
\eref{def:Pw}.

\subsection{Trajectory-to-trajectory fluctuations and the shape of $P(\omega)$}

To  better grasp  the  relation  between the  trajectory-to-trajectory
fluctuations and  the distribution  of the uniformity  index, consider
$P(\omega)$  of \eref{eq:Pw-exptrunc}  corresponding  to  the FPTD  of
\eref{eq:Pt-exptrunc}.  First, one  readily  notices that  $P(\omega)$
vanishes exponentially fast  when $\omega \to 0$ or $\omega  \to 1$ so
that  $P(\omega =  0) =  P(\omega =  1) =  0$. Second,  $P(\omega)$ is
  symmetric under  the  replacement $\omega  \to  1 -  \omega$.
Since  the distribution  $\Psi(\tau)$   possesses   moments of
arbitrary order, our first guess would be that $P(\omega)$ is always a
bell-shaped function with a maximum at $\omega = 1/2$.

\begin{figure}[t!]
  \centerline{\includegraphics*[width=1.0\textwidth]{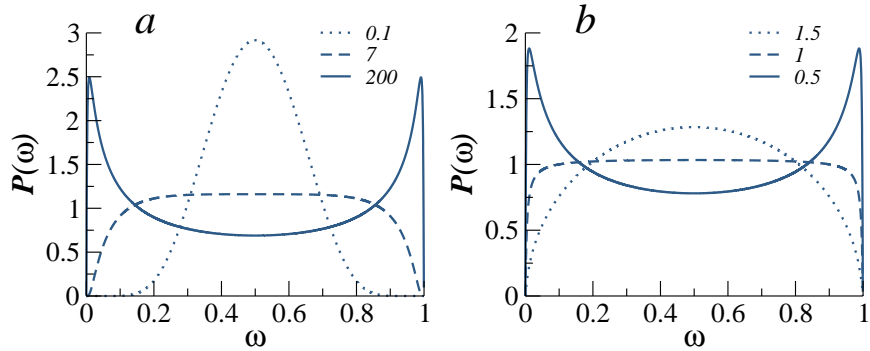}}
  \caption{The distribution $P(\omega)$ of \eref{eq:Pw-exptrunc}:
     Panel $(a)$, for $\mu=0.25$ and different values of
     $b$. Panel $(b)$, for $b=200$ and different values of
     $\mu$. Parameter $a$ is set equal to $1$.}
  \label{fig2}
\end{figure}

To study  the shape  of the distribution  of the uniformity  index, we
expand  $P(\omega)$ in  Taylor series  around its  symmetric point
$\omega   =   1/2$   up   to   second   order   in   $\left(\omega   -
  \frac{1}{2}\right)^4$,
\begin{equation}
P(\omega)\approx \frac{2 K_{2 \mu}(2 C)}{K^2_{\mu}(C)} \, \Bigg[ 1 + 
4 \left(1 - \mu - C \frac{K_{2 \mu - 1}(2 C)}{K_{2 \mu}(2 C)}\right) \left(\omega -
  \frac{1}{2}\right)^2\Bigg] \ ,
\end{equation}
where $C\equiv 2 \sqrt{a/b}$. 
Inspecting the sign of the coefficient before the quadratic term, \ie,
\begin{equation}
\label{g}
g = 1 - \mu - C \frac{K_{2 \mu - 1}(2 C)}{K_{2 \mu}(2 C)},
\end{equation}
we notice that
\begin{itemize}
\item For $\mu > 1$, $g$ is  always negative for any value of $b/a$ so
  that  here the  distribution $P(\omega)$  is a  bell-shaped function
  with a maximum at $\omega = 1/2$.
\item For  $\mu = 1$,  $g$ is negative  and approaches $0$  from below
  when  $b/a \to  \infty$. It  means that  $P(\omega)$ is  generally a
  bell-shaped function  with a maximum  at $\omega  = 1/2$, but  it
  becomes progressively flatter  when  $b/a$ is  increased, so  that
  ultimately $P(\omega) \approx  1$ apart from very  narrow regions at
  the edges for $b/a \gg 1$.
\item  For $0  \leq \mu  <  1$ there  always exists  a critical  value
  $y_c(\mu)$ of the parameter $y = b/a$ which is defined implicitly as
  the solution  of \eref{g} for  $g = 0$.   For $b/a <  y_c(\mu)$, the
  distribution  $P(\omega)$ is unimodal  with a  maximum at  $\omega =
  1/2$.  For $b/a  =  y_c(\mu)$, the  distribution  is nearly  uniform
  except for  narrow regions  in the vicinity  of the  edges. Finally,
  which is  quite surprising  in view  of the fact  that in  this case
  $\Psi(\tau)$  possesses  all  moments,  for  $b/a  >  y_c(\mu)$  the
  distribution $P(\omega)$  is bimodal with  a characteristic M-shaped
  form, two maxima  close to $0$ and $1$ and $\omega  = 1/2$ being the
  least probable value.
\end{itemize}
We depict in
  Fig.~\ref{fig2} three characteristic forms of $P(\omega)$ for $\mu =
  1/2$ and three different values of $b/a$.

Therefore,  for $\mu  \geq 1$  two independent  BMs will  most probably
reach the target  simultaneously. For $0 \leq \mu  < 1$ two distinctly
different  situations  are  possible:  if  $b/a$  is  less  than  some
well-defined critical value $y_c(\mu)$,  then most likely both BM will
arrive  to  the  target for  the  first  time  together.  If,  on  the
contrary, $b/a$ exceeds  this critical value, the event  in which both
BMs arrive  to the  location of the  target simultaneously will  be the
least probable one, yielding very large fluctuations in the MFPT
obtained from different trajectories.

In the  rest of this chapter  we present several  concrete examples in
which  the transition between unimodal and  bimodal distributions
of $P(\omega)$ is observed.

\section{One-dimensional Brownian Motions}

In this section we  study the trajectory-to-trajectory fluctuations of
the first passage time, in terms of the distribution of the uniformity
index $\omega$ for two independent identical BM.

\subsection{BM in the semi-infinite interval with a bias}

We  first  consider   a  BM   in   the  semi-infinite  interval
$\mathcal{S}\equiv[0,\infty)$ with  $\partial\mathcal{S}_a =\{0\}$ and
$\partial\mathcal{S}_r  =\{\infty\}$, in  the presence  of a  constant
bias   pointing  towards   the   target,  for   which   the  FPTD   in
\eref{eq:Pt-exptrunc} with $\mu = 1/2$ is exact.  In this case one has
$a = x_0^2/4 D$, where $x_0$  is the starting point, $D$ the diffusion
coefficient, and $b  = 4 D/v^2$ ($v$ being the  drift velocity). Hence,
the  P\'eclet  number,  defined  as  the ratio  between  the  rate  of
advection and the rate of diffusion,  is $\mathrm{Pe} = 2 \sqrt{a/b} =
x_0    |v|/2    D$    (see,    \eg,
Ref.~\refcite{sid}). Consequently, we can make a following statement:

Consider two  independent identical  BMs on a  semi-infinite interval,
starting  at  the   same  point  $x_0$,  having   the  same  diffusion
coefficient  $D$  and experiencing  the  same  bias $F$  which  points
towards the  origin so  that the drift  velocity of both  BMs is  $v <
0$. Then, an event in which both  BMs arrive for the first time to the
origin simultaneously is
\begin{itemize}
\item[---] the \textit{least} probable if $\mathrm{Pe} < \mathrm{Pe}_c$\,,
\item[---] the \textit{most} probable if $\mathrm{Pe} > \mathrm{Pe}_c$\,,
\end{itemize}
where $\mathrm{Pe}_c$ is the solution of the transcendental equation
\begin{equation}
1 = 2 \mathrm{Pe}_c \frac{K_0(2 \mathrm{Pe}_c)}{K_1(2 \mathrm{Pe}_c)} 
\approx 0.666 \ .
\end{equation}
Therefore,  the  MFPT  to  the  target,  which  in  this  case  equals
$x_0/|v|$, might  be an appropriate  measure of the  search efficiency
for sufficiently large P\'eclet numbers, but definitely is not
in   case  of   small   $\mathrm{Pe}$.   In   the   latter  case   the
trajectory-to-trajectory  fluctuations are  significant  and the  mean
value is not representative of the actual behaviour.

\subsection{BM on a finite interval}

We consider  now BMs  in the finite  interval $\mathcal{S}\equiv[0,L]$
with   $\partial\mathcal{S}_a   =\{0\}$   and   $\partial\mathcal{S}_r
=\{L\}$, that  are initiated at $x=x_0$  at time $t=0$. In  this case the
FPTD can be obtained exactly as the infinite sum
\begin{equation}
\label{fpt1}
\Psi_{x_0}(\tau) = \frac{2 \pi D}{L^2} \sum_{n = 0}^{\infty}
A_n\left(\frac{x_0}{L}\right)
 \exp\left(- \frac{\pi^2 (n + 1/2)^2 D \tau}{L^2}\right),
\end{equation}
where  the  coefficients  $A_n\left(\frac{x_0}{L}\right) =  \left(n  +
  \frac{1}{2}\right)\sin\left(\frac{\pi  (n  +  1/2)  x_0}{L}\right)$.
Consequently,  after  some  calculations the  normalised  distribution
$P(\omega)$ has the following form\cite{carlos}
\begin{equation}
\label{pw1d}
P(\omega) 
= \frac{2}{\pi} \frac{d}{d\omega} \sum_{m = 0}^{\infty}
\frac{\sin\left(\pi (m+1/2) \frac{x_0}{L}\right)}{m + 1/2} 
\frac{\cosh\left(\pi (m+1/2) \sqrt{\frac{1 - \omega}{\omega}} \left(1 - \frac{x_0}{L}\right)\right)}{\cosh\left(\pi (m+1/2) \sqrt{\frac{1 - \omega}{\omega}}\right)} \,.
\end{equation}

\begin{figure}[t!]
  \centerline{\includegraphics*[width=1.0\textwidth]{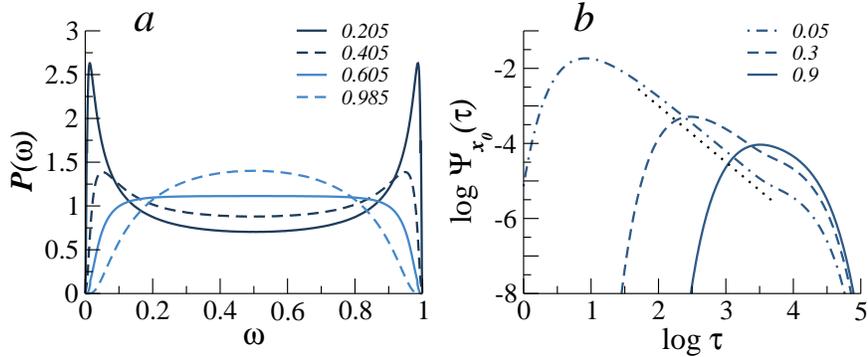}}
  \caption{Two BMs in the interval $[0,L]$.  Panel $(a)$: Distribution
    $P(\omega)$   in   Eq.~(\ref{pw1d})   for  different   values   of
    $x_0/L$. Panel $(b)$:  FPTD $\Psi_{x_0}(\tau)$ in Eq.~(\ref{fpt1})
    for $L= 100$, $D = 1/2$ and different values of $x_0/L$. The
    dotted curve corresponds to a decay $\sim\tau^{-3/2}$.
    \label{1dd}}
\end{figure}
Panel $(a)$  of Fig.~\ref{1dd}  shows the distribution  $P(\omega)$ in
\eref{pw1d} for different values  of the ratio $x_0/L$, evidencing the
transition in the shape of  $P(\omega)$ at the critical value $\approx
0.605$. 

Furthermore,  the  moments  of  arbitrary  order  of  $\Psi(\tau)$  in
Eq.~(\ref{fpt1}) can be straightforwardly calculated. One obtains
\begin{equation}
\big<\tau^m\big> \equiv \int^{\infty}_0 \tau^m \, d\tau \, \Psi(\tau) = \frac{(-1)^m \sqrt{\pi}}{\Gamma(m + 1/2)} \left(\frac{ L^2}{D}\right)^m E_{2 m}\left(\frac{x_0}{2 L}\right) \,,
\end{equation}
where $E_{2 k}(\cdot)$  are the Euler polynomials.  Consequently, the mean
and the variance are given by
\begin{equation}
\big<\tau\big> = - 2 \frac{ L^2}{D} \, E_2\left(\frac{x_0}{2 L}\right),
\end{equation}
and
\begin{equation}
k_2 = \big<\tau^2\big> - \big<\tau\big>^2 = \frac{4}{3} \frac{L^4}{D^2} \left(E_4\left(\frac{x_0}{2 L}\right) - 3 E_2^2\left(\frac{x_0}{2 L}\right)\right) \,
\end{equation}
respectively. One can readily check that both statistical measures are
monotonically increasing functions of $x_0/L$ and do not show any sign
of a particular behaviour at  $x_0/L \approx 0.605$. The same occurs
for   higher  order   cumulants  of   the   distribution  $\Psi(\tau)$
\cite{carlos}.

This puts  in evidence the  superior sensitivity of the  statistics of
the uniformity index $\omega$.  As we show in sections~\ref{sec:2dNET}
and  \ref{sec:NET} ,  the  unimodal or  bimodal  shape of  $P(\omega)$
classifies the parameter  space into regions inside which  the MFPT is
either \emph{meaningful} or \emph{meaningless}.

\subsection{Arcsine law mechanism}
\label{sec:arcsine}

We  found that  any  two BMs  arrive to  the  target at  progressively
distinct  times the  closer they  are initially  to its  location, and
should  most probably  arrive  together  when they  are  far from  it.
Notwithstanding this  counterintuitive result, such a  behavior is the
same as the one behind the  famous arcsine law for the distribution of
the fraction of time spent by  a random walker on a positive half-axis
\cite{arcsine}: \vspace{2mm}

\emph{Once one of the BMs goes  away from the target, it finds it more
  difficult to return than to keep on going away}.
\vspace{3mm}

How does  this relate  to the  properties of  the distribution  of the
first  passage time?   In  terms of  the FPTD  of  \eref{fpt1} in  the
previous  section, $x_0$  ($a^{1/2}$) and  $l$ ($b^{1/2}$)  define the
effective  size of  the  region in  which  the decay  of  the FPTD  is
governed by the  intermediate power-law tail.  The  larger this region
is   the  larger   the  fluctuations   in  the   first  passage   time
become\cite{carlos}.    This    is   shown    in   panel    $(b)$   of
Fig.~\ref{1dd}.  All three  curves  show an  exponential behavior  for
small and  large values  of $\tau$,  as in  \eref{eq:Pt-exptrunc}. For
large  $\tau$ all  three curves  merge  which signifies  that at  such
values of  $\tau$ the characteristic  decay time is dependent  only on
$l$. The lower cut-off is clearly dependent only on the starting point
$x_0$. Evidently the  power-law region characterized by  a decay $\sim
t^{-3/2}$,  grows with  decreasing $x_0/l$,  \ie, when  the BMs  start
closer to the target.

\section{Two-dimensional Brownian Motions}

In this section we analyse the statistics of $\omega$ for BM in
different two-dimensional bounded domains. 

\subsection{BM in a disc with a reflecting boundary}

We first consider BM in a disc  of radius $L$, centered at the origin
$\mathcal{S}=\{\mathbf{r}    :     |\mathbf{r}|    <     L\}$,    with
$\partial\mathcal{S}_a  =\{\mathbf{r}   :  |\mathbf{r}|  =   l\}$  and
$\partial\mathcal{S}_r =\{\mathbf{r}  : |\mathbf{r}| = L\}$, where $\mathbf{r} \in \mathbb{R}^2$ and $l <
L$.  Therefore,  the target  consist in a  concentric disc  of smaller
radius $l$.

Suppose next that a BM starts at some point at distance $r_0$ from the
origin and hits  the target for the first time  at time moment $\tau$.
The FPTD $\Psi_{r_0}(\tau)$ is explicitly given by \cite{carlos}
\begin{equation}
\label{fpt2}
\Psi_{r_0}(\tau) = \frac{l D}{Z} \sum_{n = 0}^{\infty} A_n(l,r_0,L)  \exp\left(- \lambda_n^2 D \tau\right),
\end{equation}
where $Z$ is the normalization,
\begin{equation*}
A_n(l,r_0,L) = \frac{U_0(\lambda_n r_0) U_0'(\lambda_n l)}{\lambda_n^2 L^2 U_0^2(\lambda_n L) - l^2 \left(U_0'(\lambda_n l)\right)^2},
\end{equation*}
\begin{equation*}
U_0(x) = Y_0(\lambda_n l) J_0(x) - J_0(\lambda_n l) Y_0(x); \quad U_0'(\lambda_n l) = \frac{d U_0(\lambda_n x)}{dx}\bigg|_{x=l}
\end{equation*}
while $\lambda_n$ are the roots of the function $Y_0(\lambda_n l)
J_1(\lambda_n L) - J_0(\lambda_n l) Y_1(\lambda_n L)$, 
arranged in an ascending  order, and $Y_n(\cdot)$ are Bessel functions
of the second kind. Note that $\lambda_n$ depends on $L$ and $l$.

In this case the distribution of the uniformity index $\omega$ can be
written as \cite{carlos}
\begin{equation}
 \label{pw2}
P(\omega) = \frac{l}{Z} \frac{d}{d\omega} \sum_{m=0}^{\infty} \frac{A_m(l,r_0,L)}{\lambda_m^2} \, \Phi\left(\lambda = \frac{1 - \omega}{\omega} D \lambda_m^2\right) \,,
 \end{equation}
 where  $\Phi(\lambda)$ is  the characteristic  function of  the first
 passage   time   distribution    in   Eq.~(\ref{fpt2})   defined   by
 \cite{sid}
\begin{equation}
\Phi(\lambda) = \frac{I_0\left(\sqrt{\frac{\lambda}{D}} r_0\right) K_1\left(\sqrt{\frac{\lambda}{D}} L\right)   + K_0\left(\sqrt{\frac{\lambda}{D}} r_0\right) I_1\left(\sqrt{\frac{\lambda}{D}} L\right)}{I_0\left(\sqrt{\frac{\lambda}{D}} l\right) K_1\left(\sqrt{\frac{\lambda}{D}} L\right)   + K_0\left(\sqrt{\frac{\lambda}{D}} l\right) I_1\left(\sqrt{\frac{\lambda}{D}} L\right)} \,.
\end{equation}

\begin{figure}[t!]
  \centerline{\includegraphics*[width=1.0\textwidth]{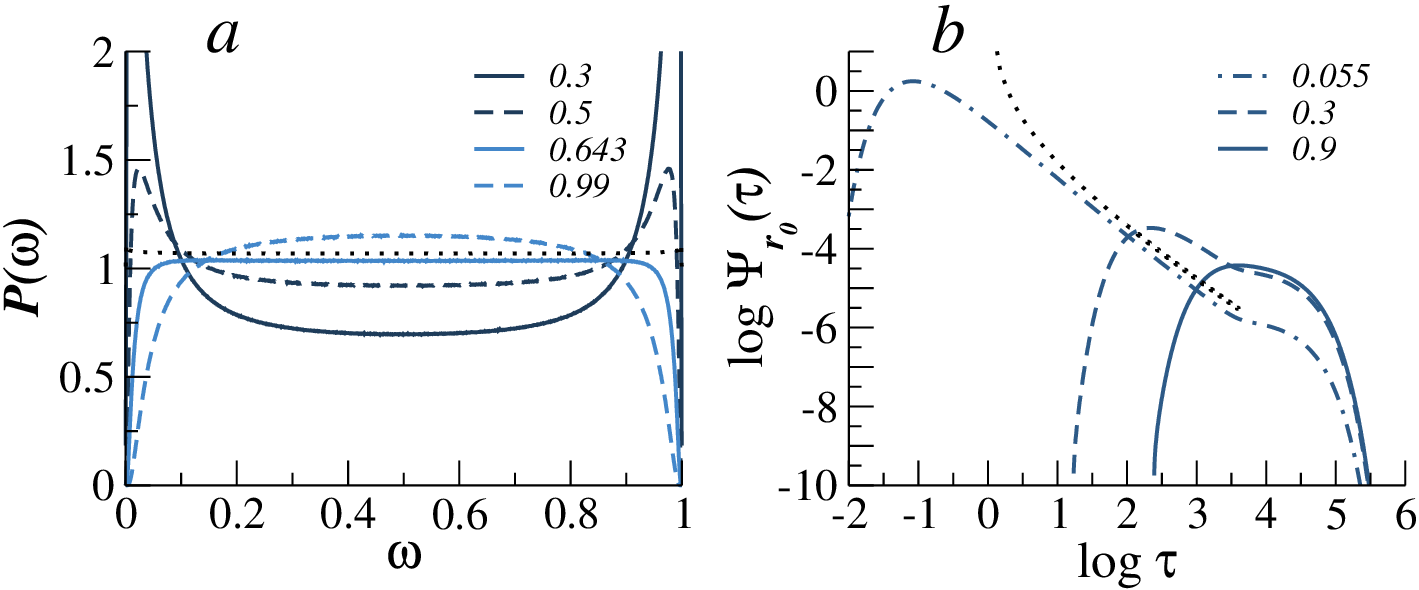}}
  \caption{Two BMs in the  two-dimensional disc of radius $L=100$ with
    a  reflecting boundary  and  concentric absorbing  disc of  radius
    $l=5$,  and  $D=1/2$.  Panel  $(a)$:  Distribution $P(\omega)$  in
    Eq.~(\ref{pw2}) for different values  of $r_0/L$. The dotted curve
    corresponds   to  $P_{\mathrm{av}}(\omega)$.   Panel   $(b)$:  FPTD
    $\Psi_{r_0}(\tau)$  in Eq.~(\ref{fpt2})  for  different values  of
    $r_0/L$.     The  dotted   curve    corresponds   to    a   decay
    $\sim1/\tau\log^2\tau$.}
  \label{2dd}
\end{figure}

Fig.~\ref{2dd} shows  the FPTD in Eq.~(\ref{fpt2})  and $P(\omega)$ in
Eq.~(\ref{pw2})  for fixed  $L$, $D$  and $r$,  and several  values of
$r_0$.   Here   we  observe  the   same  trend  as  in   the  previous
one-dimensional  example, namely  the  FPTD becomes  broader when  the
starting point  of the BMs is closer  to $\partial\mathcal{S}_a$, with
an intermediate power-law behavior with a logarithmic correction $\sim
1/\tau \log^2(\tau)$ (see the dashed curve in Fig.~\ref{2dd}-$b$). The
critical  value   for  the  transition  between   bell-  and  M-shaped
$P(\omega)$ is numerically found to be $r_0/L \sim 0.643$. Moreover, if
the  starting   point  of   two  BM  is   averaged  over   the  domain
$\mathcal{S}$,  one   readily  finds,   by  averaging  the   result  in
Eq.~(\ref{pw2}), that $P_{av}(\omega) \equiv  1$ (see the dashed curve
in Fig.~\ref{2dd}-$b$).

\subsection{BM in a disc with aperture}
\label{sec:2dNET}

For  other simple geometries  but mixed  absorbing/reflecting boundary
conditions, it is not possible  in general to obtain analytical closed
expressions for the FPTD.   However, the examples discussed previously
suggest  that the arcsine  law mechanism  in section~\ref{sec:arcsine},
underlying   the    appearance   of   large   trajectory-to-trajectory
fluctuations of the  first passage time, is generic  in bounded domains
of arbitrary shape.

For instance, consider a circular  domain of unit radius (in polar
coordinates $\mathbf{r} = (r,\theta)$), $\mathcal{S}=\{\mathbf{r} : r <
1\}$,  and the following  boundary conditions: the  segment with
$|\theta|<\Theta/2$ is absorbing while the remaining part of the outer
circle is reflective.  The aperture of  the circular domain is thus an
arc of length $\Theta$.

Obtaining an  expression for the FPTD involves  the cumbersome problem
of  finding  the  solution   of  the  diffusion  equation  with  mixed
Dirichlet-Neumann boundary conditions.  Here we content ourselves with
the  numerical analysis  of  the  problem.  Let  each  BM commence  at
position $(\rho_0,\theta_0)$ inside the  unit circle and determine the
set  $\{\tau_i\}$  of first  passage  times  to  the location  of  the
aperture.  From  these data one obtains $P(\omega)$.   To easily grasp
the geometric  dependence of the  first passage time  fluctuations, we
considered  the parameter  $\chi$  that quantifies  the  shape of  the
distribution  of  $\omega$\cite{thiago}.   Performing  a  fit  of  the
numerically obtained $P(\omega)$ to a quadratic polynomial of $\omega$
in the domain $0.05<\omega<0.95$,  we define $\chi$ as the coefficient
of the quadratic term, so that the sign of $\chi$ determines the shape
of  $P(\omega)$:  $\chi<0$ corresponds  to  the unimodal,  bell-shaped
distribution,  $\chi>0$ signifies  that the  distribution  is bimodal,
M-shaped,  and a  zero value  of  $\chi=0$ means  that $P(\omega)$  is
uniform.

\begin{figure}[t!]
\centerline{\includegraphics*[width=0.6\linewidth]{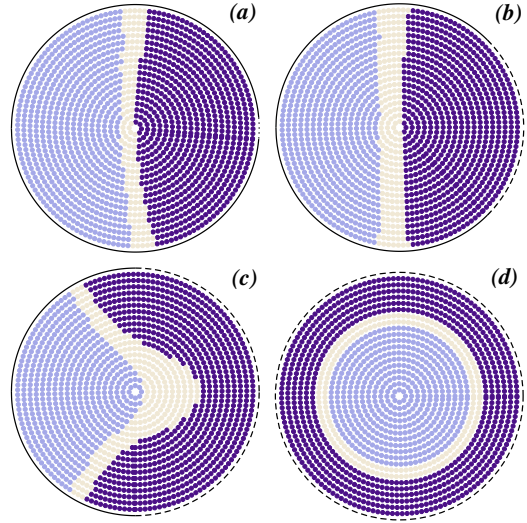}}
\caption{Phase-chart  for  the shape  of  the uniformity  distribution
  $P(\omega)$  for a  BM in  the  unit disc  with reflective  boundary
  (solid line) and  an aperture of size $\Theta$  corresponding to the
  absorbing   boundary  (dashed   line).   (a)   $\Theta=\pi/18$,  (b)
  $\Theta=\pi/2$,  (c) $\Theta=\pi$,  and (d)  $\Theta=2\pi$. Starting
  locations are  coloured light blue if  $\chi<-\chi_\star$, dark blue
  if  $\chi>\chi_\star$  and  beige  if  $|\chi|<\chi_  \star$,  where
  $\chi_\star=1$.}
\label{fig:circle}
\end{figure}
 
In  Fig.~\ref{fig:circle} we  show the  phase-chart for  the shape  of
$P(\omega)$ for  different sizes  of the aperture  \footnote{Note that
  the   case  shown   in  Fig.~\ref{fig:circle}   (d)  reduces   to  a
  one-dimensional   problem\cite{carlos}.}.    As  in   all   previous
examples,  the closer  the  starting  position of  the  BM  is to  the
absorbing boundary  the larger the  fluctuations of the  first passage
time become.  One thus expects  that very near the  absorbing boundary
the standard deviation  of the first passage time  becomes much larger
than its mean.  We note that  this result is generic irrespectively of
the aperture size.

The same analysis was recently carried out for other different shapes,
such  as pie-wedges  or triangles  in Ref.~\refcite{thiago},  with the
same conclusions.

\section{Three-dimensional Brownian Motions}

\subsection{BM in a sphere with a reflecting boundary}

Consider  a BM in  a  sphere  of radius  $L$  (in spherical  coordinates
$\mathbf{r} = (\rho,\theta,\phi)$), $\mathcal{S}=\{\mathbf{r} : \rho <
L\}$ with  reflecting boundary $\partial\mathcal{S}_r  =\{\mathbf{r} :
\rho =  L\}$, and  a concentric  absorbing sphere of  radius $l  < L$,
$\partial\mathcal{S}_a =\{\mathbf{r} : \rho = l\}$.

For this geometry,  the distribution of the first  passage time $\tau$
of a BM starting at a distance $\rho_0$ from the origin to the surface of
the target is given explicitly by\cite{carlos}
\begin{equation}
\label{fpt3}
\Psi_{\rho_0}(\tau) = \frac{D}{Z} \sum_{n=0}^\infty A_n(l,\rho_0,L)
\exp\left(-\lambda_n^2D\tau\right) \ ,
\end{equation}
where the coefficients
\begin{equation*}
A_n(l,\rho_0,L) =  \frac{2 u_0(\lambda_n \rho_0) u_0^\prime(\lambda_n
l)}{G(l,L,\lambda_n)} \ ,
\end{equation*}
with $u_0(x)   =   y_0(\lambda_n   l)   j_0(x)   -
j_0(\lambda_n  l) y_0(x)$,  $u_0^\prime(\lambda_n l)  = du_0(\lambda_n
x)/dx|_{x=l}$, and
\begin{eqnarray*}
G(l,L,\lambda_n) &=& Rj_0(\lambda_nL)
u_0(\lambda_nL)(\lambda_nLj_0(\lambda_nl) - y_0(\lambda_nl))
- \nonumber\\
&-& \frac{j_0(\lambda_nl)y_0(\lambda_nl)}{\lambda_n}
 + \frac{L-l}{l^2\lambda_n^4} \ .
\end{eqnarray*}
The  set   $\{\lambda_n\}$  are  the  roots  of   $  y_0(\lambda_n  l)
j_1(\lambda_n L) - j_0(\lambda_n l) y_1(\lambda_n L) $, arranged in an
ascending order, while $j_n(\cdot)$ and $y_n(\cdot)$ are the spherical
Bessel functions of the first and of the second kind, respectively.

\begin{figure}[t!]
  \centerline{\includegraphics*[width=1.0\textwidth]{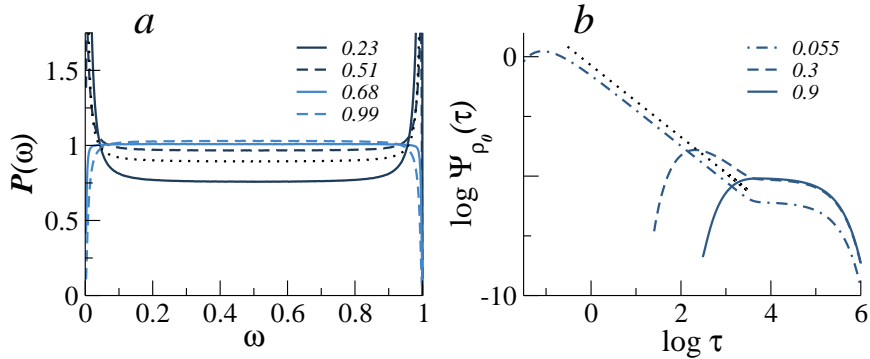}}
  \caption{BM in a sphere of radius $L=100$ with
    a  reflecting boundary  and  concentric absorbing sphere of  radius
    $l=5$,  and  $D=1/2$.  Panel  $(a)$:  Distribution $P(\omega)$  in
    Eq.~(\ref{pw3}) for different values  of $\rho_0/L$. The dotted curve
    corresponds   to  $P_{\mathrm{av}}(\omega)$.   Panel   $(b)$:  FPTD
    $\Psi_{\rho_0}(\tau)$  in Eq.~(\ref{fpt3})  for  different values  of
    $\rho_0/L$.     The   dotted   curve    corresponds   to    a   decay
    $\tau^{-3/2}$.}
  \label{3dd}
\end{figure}

The  corresponding distribution  $P(\omega)$ for  two  independent BMs
starting at a distance $\rho_0$ from the origin is \cite{carlos}
\begin{eqnarray}
\label{pw3}
P(\omega) &=& \frac{1}{Z^2} \sum_{n,m=0}^\infty
\frac{A_n(l,\rho_0,L) A_m(l,\rho_0,L)}{\left(\omega \lambda_n^2 + (1 - \omega) \lambda_m^2\right)^2} = \nonumber\\
&=& \frac{1}{Z} \sum_{m=0}^\infty \frac{A_m(l,\rho_0,L)}{\lambda_m^2} \, \Phi\left(\lambda = \frac{1 - \omega}{\omega} D \lambda_m^2\right) \,,
\end{eqnarray}
where $\Phi(\lambda)$  is the characteristic  function of $\Psi(\tau)$
in Eq.~(\ref{fpt3}), defined by \cite{sid}
\begin{equation}
\Phi(\lambda) = \frac{l}{\rho_0} \frac{\sinh\left(\sqrt{\frac{\lambda}{D}} \left(L - \rho_0\right)\right) - \sqrt{\frac{\lambda}{D}} \, R \, \cosh\left(\sqrt{\frac{\lambda}{D}} \left(L - \rho_0\right)\right)}{ \sinh\left(\sqrt{\frac{\lambda}{D}} \left(L - l\right)\right) - \sqrt{\frac{\lambda}{D}} \, R \, \cosh\left(\sqrt{\frac{\lambda}{D}} \left(L - l\right)\right)} \,.
\end{equation}

The distributions  in Eqs.~(\ref{fpt3})  and (\ref{pw3}) are  shown in
Fig.~\ref{3dd}  for fixed  $L$, $l$  and  $D$, and  several values  of
$\rho_0$. An intermediate power-law $\sim \tau^{-3/2}$ is apparent for
$\rho_0/L=0.055$, persists  for one  decade for $\rho_0/L=0.3$  and is
entirely absent for  $\rho_0/L=0.9$, \ie, when the  BM initiates close
to the reflecting  boundary (see Fig.~\ref{3dd}-$b$).  Correspondingly
we find  that $P(\omega)$  has a different  modality depending  on the
value of the ratio $\rho_0/L$ with  a critical value of $\rho_0/L \sim
0.68$  (see   Fig.~\ref{3dd}-$a$).   For  $\rho_0/L  >   0.68$,  the
distribution has, in  principle, a maximum at $\omega =  1/2$ but this
maximum is  almost invisible so  that visually the  distribution looks
more like  a uniform  one, as  compared to  the 1D  case in  which the
maximum  is  more  apparent.   When the  starting  point  $\rho_0$  is
uniformly distributed in $\mathcal{S}$,  we again find $P_{av}(\omega)
\equiv 1$ (dashed curve in Fig.~\ref{3dd}-$a$).

\subsection{Narrow Escape Problem}
\label{sec:NET}

As  a last example  we consider  the problem  of the  so-called Narrow
Escape Time  (NET).  This  problem, generic in  cellular biochemistry,
consists of estimating  the time that a randomly  moving particle spends
inside a bounded domain before it escapes through a narrow aperture on
the boundary of the domain. The particle  can be an ion, a ligand or a
molecule diffusing  inside a cell,  a microvesicle, a  compartment, an
endosome,  a  caveola, a  dendritic  spine,  etc.,  and a  variety  of
processes in which the importance  of the NET problem is striking have
been discussed in the past \cite{bioNET1,bioNET2,bioNET3,NEP4,NEP5}. 

\begin{figure}[t!]
  \centerline{\includegraphics*[width=0.65\textwidth]{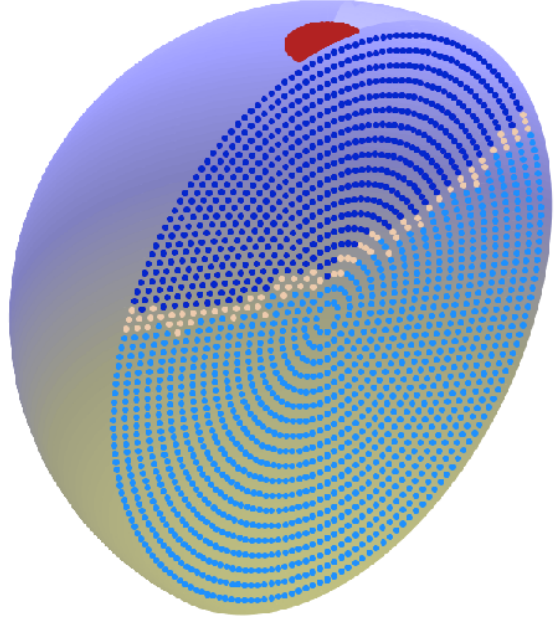}}
  \caption{Cut of the phase-chart for the shape of $P(\omega)$ for the
    NET     problem      with     $p_a=0$,     $\sigma=1/4\pi$     and
    $D_V=0.000416=D_S=0.000625$.  Starting locations are coloured light
    blue  if $\chi<-\chi_\star$, dark  blue if  $\chi>\chi_\star$ and
    beige  if  $|\chi|<\chi_ \star$,  where  $\chi_\star=0.1$.  The  red
    region  on  the  sphere  indicates  the  position  of  the  narrow
    aperture. The  phase-chart has  rotation symmetry with  respect to
    the position of the aperture.}
  \label{fig:NETP}
\end{figure}
 
In section~\ref{sec:2dNET}  we discussed a  two-dimensional version of
this problem.  In  three dimensions, for a spherical  domain of radius
$L$,  reflecting everywhere except  at a  narrow circular  aperture of
radius $\sigma\ll L$ in which the boundary is absorbing, the survival
probability  at  large  times is  $\mathscr{S}_{\mathbf{r}_0}(t)  \sim
\exp(-t/\tau_{3D})$, where  the characteristic time  decay $\tau_{3D} =
\pi L^3/3D\sigma$ corresponds to the MFPT\cite{purcell}.

Here we  discuss the  statistics of the  uniformity index for  the NET
problem.  Consider a sphere of unit radius $\mathcal{S}=\{\mathbf{r} :
\rho    <    1\}$   (in    spherical    coordinates   $\mathbf{r}    =
(\rho,\theta,\phi)$), and  the following boundary  conditions: a solid
angle $|\Omega|<\sigma$  is absorbing while the remaining  part of the
sphere is reflective.  Furthermore, consider a BM inside $\mathcal{S}$
with  a   diffusion  coefficient  $D_V$,  and   initially  located  at
$(\rho_0,\theta_0,\phi_0)$.   When  the  Brownian  particle  hits  the
sphere   $\partial\mathcal{S}$  it   becomes   weakly  absorbed   with
probability $p_a$  and starts  to diffuse, with  diffusion coefficient
$D_S$, along the surface of the sphere. During a time interval $\Delta
t$, the particle  detaches from the sphere with  probability $p_a \Delta t$,
and continue its  motions until it hits the narrow  aperture at  time
$\tau$.

Considering  couples  of  BMs   one  obtains  $P(\omega)$  as  before.
Fig.~\ref{fig:NETP}   shows  the   phase-chart   for   the  shape   of
$P(\omega)$, as defined in  section~\ref{sec:2dNET}, for $p_a=0$, \ie,
when  the  BM  never  gets   attached  to  the  sphere.   The  results
qualitatively resemble  those obtained in  section~\ref{sec:2dNET} for
the    two-dimensional     disc    for    narrow     apertures    (see
Fig.~\ref{fig:circle}-$a$).   Note that  the statistics  of the  first
passage time inherit  the polar symmetry of the  geometry with respect
to the position of the aperture.

Our results  indicate that for BMs  starting near the  location of the
aperture, the MFPT  has little significance to determine  the NET.  We
have also considered finite values of the attachment probability $p_a$
(finite  particle-surface affinity),  obtaining  qualitatively similar
results than the ones shown in Fig.~\ref{fig:NETP} for $p_a=0$.

\section{Conclusions}

In this Chapter we discussed a novel diagnostic method which allows to
quantify  straightforwardly  and   meaningfully  the  effects  of  the
trajectory-to-trajectory  fluctuations   in  first  passage  phenomena
characterised  by  ``narrow''  distributions  $\Psi(\tau)$  possessing
moments  of  arbitrary  order.    This  diagnostic  is  based  on  the
simultaneity  concept of the  first passage  events. To  this purpose,
instead  of the  original  random variable  $\tau$ characterising  the
first   passage   event,   we   introduced  a   variable   $\omega   =
\tau_1/(\tau_1+\tau_2)$,  where $\tau_1$  and  $\tau_2$ are  identical
independent    random   variables    with   the    same   distribution
$\Psi(\tau)$. Physically,  the uniformity index $\omega$  shows us the
likelihood of the event that  two different trajectories arrive to the
prescribed location for the first time simultaneously.

We  demonstrated  that the  characteristic  shapes  of the  associated
distribution of  the uniformity index  $\omega$ are very  sensitive to
the  trajectory-to-trajectory fluctuation.  In  some cases,  when such
fluctuations have only   little effect so that the  MFPT can be considered
as  a meaningful  property, the  distribution $P(\omega)$  is unimodal
with a maximum  at $\omega = 1/2$, such  that $\tau_1 = \tau_2$
is the  most probable event. In  other cases, $P(\omega)$  is shown to
have a bimodal form with two maxima close to $\omega =0$ and $\omega =
1$ and  a local minimum  at $\omega =  1/2$.  This means that  in such
situations  two different  first passage  events will  be  most likely
characterised  by very different  $\tau_1$ and  $\tau_2$, so  that the
trajectory-to-trajectory fluctuations  will be very  important and the
MFPT cannot be considered as a meaningful characteristic measure of the
process.   In  this regard,  the  underlying ``narrow''  distributions
$\Psi(\tau)$  may show  a  behaviour specific  to broad  distributions
which do not possess all moments.

To illustrate this concept, we considered several examples of Brownian
motion in  bounded domains, in  different spatial dimensions  and with
different locations of the  target.  We demonstrated that, remarkably,
within the same domain the  very shape of the distribution $P(\omega)$
depends crucially  on the location of  the target and  on the starting
point of the Brownian motion  --- trajectories starting from one point
may  exhibit  considerable  fluctuations  with respect  to  the  first
passage  to the target,  while trajectories  starting from  some other
point may  arrive to the  target location almost  simultaneously.  For
some   geometries,  we   evaluated  the   ``phase  charts''   for  the
distribution $P(\omega)$  showing which shape ---  unimodal or bimodal
--- this  distribution will have  for different starting  point within
each domain.

We  expect  that a  similar  behaviour  will  be observed  in  bounded
one-dimensional  geometries  for fractional  BM  with arbitrary  Hurst
index $H$ or for $\alpha$-stable L\'evy flights with $0<\alpha<1$, and
more  generally, for bounded  geometries with  compact or  non compact
exploration.

\bibliographystyle{ws-rv-van}
\bibliography{FPT-book}

%\printindex[aindx]                 % to print author index

\end{document}